\documentclass[aps,preprint,showpacs,showkeys]{revtex4}%
\usepackage{amsfonts}
\usepackage{amsmath}
\usepackage{amssymb}
\usepackage{graphicx}%
\begin{document}
\title{Analysis of radiation-pressure induced mechanical oscillation of an optical
microcavity }
\author{T.J. Kippenberg, H. Rokhsari, T. Carmon, A. Scherer and K.J. Vahala}
\email{vahala@its.caltech.edu}
\affiliation{Thomas J. Watson Laboratory of Applied Physics, California Institute of Technology}

\begin{abstract}
The theoretical work of V.B. Braginsky predicted that radiation pressure can
couple the mechanical, mirror-eigenmodes of a Fabry-P\'{e}rot resonator to
it's optical modes, leading to a parametric oscillation instability. This
regime is characterized by regenerative mechanical oscillation of the
mechanical mirror eigenmodes. We have recently observed the excitation of
mechanical modes in an ultra-high-Q optical microcavity. Here, we present a
detailed experimental analysis of this effect and demonstrate that radiation
pressure is the excitation mechanism of the observed mechanical oscillations.

\end{abstract}
\pacs{PACS number: 42.65.Sf, 42.65.Ky, 42.65.Yj.}
\maketitle

The work of V.B. Braginsky\cite{Braginsky4,Braginsky5} predicted that pressure
induced by circulating radiation in a Fabry-P\'{e}rot resonator can couple the
optical modes to the mechanical mirror eigenmodes. The coupling can lead to a
parametric oscillation instability, characterized by regenerative oscillation
of the mechanical mirror eigen-modes. This mechanism has been studied
theoretically for its possible role in setting a detection sensitivity limit
in the laser interferometer gravitational wave observatory (LIGO)
\cite{Abramovici, Amelin}, but has so far not been observed experimentally.

Recently, we have observed a nonlinear mechanism\cite{Rokhsari} in
ultra-high-Q toroid microcavities\cite{Armani} that is distinct from other
nonlinear mechanisms already observed in these
structures\cite{Kippenberg4,Spillane}. The geometry for observation of this
nonlinearity is a standard one in which a wave (here referred to as the pump)
is coupled from a waveguide to a microcavity mode. The nonlinearity manifests
itself as oscillations in the pump power transmitted past the micro-cavity.
These oscillations are observed to occur at a distinct threshold pump power
level and have spectral components at characteristic frequencies. Numerical
modeling and spectral analysis reported in Ref. \cite{Rokhsari} revealed that
the observed oscillations are due to regenerative oscillation of certain
mechanical eigenmodes of the toroid microcavity. In this letter, we
demonstrate that the observed mechanical oscillations are caused by radiation
pressure, and specifically rule out another mechanism (thermal
effects\cite{Zalalutdinov2} ). As such, this work confirms the first
observation of radiation-pressure-induced parametric oscillation instability
as predicted by Braginsky.

The theoretical treatment of Braginsky \cite{Braginsky4} considered mechanical
oscillations of Fabry-P\'{e}rot mirror eigenmodes which leads to Stokes and
anti-Stokes sidebands (at frequencies, $\omega_{0}\pm\omega_{m} $ where
$\omega_{0}$ is the optical and $\omega_{m}$ the mechanical frequency). It was
shown\cite{Braginsky4} that if the Stokes field coincides with an adjacent
optical cavity mode the phenomenon of parametric oscillation instability can
occur. In contrast to the Braginsky theory, we observed mechanical
oscillations of \textit{several} mechanical modes (above a certain threshold)
when the mechanical resonance frequencies ($\omega_{m}$) produce Stokes and
anti-Stokes fields that fall \textit{within} the \textit{same} cavity
resonance (i.e. $\omega_{m}<\frac{\omega_{0}}{Q}$)\cite{Rokhsari}. For
Q-factors in the range of 10$^{6}$-10$^{8}$ this corresponds to frequencies in
the range of ca. 1-100 MHz which coincides with the range of the first three
fundamental mechanical modes of the toroid microcavities employed in this
work. Fig. 1a shows the first, three mechanical modes of a toroid microcavity
and Fig. 1b their frequency dependence on cavity length. Note that the mechanical motion
causes modulation of the optical pathlength of the toroid cavity modes,
causing the excitation of optical sidebands.\ These fields appear in the
cavity transmission spectrum, as shown in Fig.1c.

To account for this scenario we have extended the coupled mode analysis of
Braginsky to the present case of Stokes and anti-Stokes frequency pairs
falling within the \textit{same} cavity resonance (For simplicity only one
pair is considered here). In addition optical coupling effects associated with
the waveguide-resonator junction are, by necessity, included in the analysis.
Using the rotating wave and the slowly varying envelope approximation for all
field amplitudes, the mutual coupling of the pump $(a_{p}),$ Stokes $(a_{S})$
anti-Stokes, $(a_{AS})$ and mechanical mode $(x_{m})$, can be described by the
following coupled-mode equations:%
\begin{align}
\frac{\partial x_{m}}{\partial t} &  =-\frac{1}{2\tau_{m}}x_{m}+\frac
{-iK_{om}}{2\sqrt{m_{eff}}C(\Gamma)}(a_{p}^{\ast}a_{AS}+a_{p}a_{S}^{\ast})\\
\frac{\partial a_{p}}{\partial t} &  =-\frac{a_{p}}{2\tau}+i\Delta\omega
a_{p}+\frac{iK_{mo}}{\sqrt{m_{eff}}\omega_{m}}(x_{m}^{\ast}a_{AS}+x_{m}%
a_{S})+\kappa s\nonumber\\
\frac{\partial a_{S}^{\ast}}{\partial t} &  =-\frac{1}{2\tau}a_{S}^{\ast
}+i\left(  \Delta\omega-\omega_{m}\right)  a_{S}^{\ast}-\frac{iK_{mo}}%
{\sqrt{m_{eff}}\omega_{m}}x_{m}a_{p}^{\ast}\nonumber\\
\frac{\partial a_{AS}}{\partial t} &  =-\frac{1}{2\tau}a_{AS}+i\left(
\Delta\omega+\omega_{m}\right)  a_{AS}+\frac{iK_{mo}}{\sqrt{m_{eff}}\omega
_{m}}x_{m}a_{p}\nonumber
\end{align}
In these equations, the optical pump is detuned from the cavity-mode
line-center by $\Delta\omega=\omega-\omega_{0}$. The Stokes and anti-Stokes
frequencies lie within the resonance bandwidth of the pump mode, and,
correspondingly, are detuned by $\Delta\omega_{AS}=\left(  \Delta\omega
+\omega_{m}\right)  $ and $\Delta\omega_{S}=\left(  \Delta\omega-\omega
_{m}\right)  $ . The first equation describes the mechanical eigenmode where
$\left\vert x_{m}\right\vert ^{2}$ is normalized to mechanical energy, i.e.
$E=\sum_{i}\int\epsilon_{i}\sigma_{i}dV$, (where $\epsilon_{i}$ and
$\sigma_{i}$ are the diagonal components of the strain and stress tensor)
which decays with the lifetime $\tau_{m}(Q_{m}=\omega_{m}\tau_{m})$.
Correspondingly, $\left\vert a_{p}\right\vert ^{2}$ is the energy in the pump
mode, $|s|^{2}$ is the launched pump power in the waveguide. The total
lifetime of the optical modes is given by $\frac{1}{\tau}=\frac{1}{\tau_{0}%
}+\frac{1}{\tau_{ex}},$where the external lifetime $(\tau_{ex})$ describes
coupling of the microcavity mode to the waveguide via $\kappa=i\sqrt{\frac
{1}{\tau_{ex}}}$ and $K\equiv\frac{\tau_{0}}{\tau_{ex}}\,$\ is the normalized
coupling constant. $C$($\Gamma)$ is a correction factor\ [$1..2$] due to
reduction of circulating power in the presence of modal
coupling\cite{Kippenberg}.\ $K_{mo}\equiv\frac{\omega_{0}}{R}$ describes the
mechanically-induced displacement of the optical cavity resonant frequency and
contains, in general, a contribution from direct spatial change as well as
refractive index changes (stress-optical effect) \cite{StressOpticalEffect}%
\cite{Shelby}. The effective coupling of optical mode to the mechanical mode
is governed by $K_{om}\equiv\frac{1}{Rn_{eff}}$ in the case of radiation
pressure\cite{Braginsky4}. The effective mass $m_{eff}$ appearing in Eqn. (1)
is calculated numerically, by evaluating the total mechanical energy\ $E_{m}$
in the mechanical mode and the corresponding harmonic radial displacement
(amplitude $r$) of the toroid periphery wherein the optical mode circulates
(compare fig. 4b)\cite{MechModulation}. Solving the coupled mode equations in
steady-state the pump-power threshold for onset of mechanical oscillations is
given by:%
\begin{align}
P_{thresh} &  =\left(  \frac{\omega_{m}}{\omega_{0}}\right)  R^{2}m_{eff}%
\frac{1}{\tau_{m}}\cdot\frac{1}{\tau_{0}}\frac{|1+K+2i\Delta\omega\tau
_{0}|^{2}}{4K}\label{EqThreshold}\\
&  \cdot\frac{1}{2\tau}[\frac{1}{1+4\tau^{2}\Delta\omega_{AS}^{2}}-\frac
{1}{1+4\tau^{2}\Delta\omega_{S}^{2}}]^{-1}\nonumber
\end{align}
Careful inspection of the last term of the threshold equation shows that
mechanical gain is possible (i.e., positive threshold power) for
$\omega>\omega_{0}($i.e. $\Delta\omega>0)$. For $\Delta\omega<0$, the
mechanical mode is damped. The need to overcome mechanical loss leads to the
$\frac{1}{\tau_{m}}-$dependence, while the dependence of radiation pressure
upon circulating optical power leads to the $\frac{1}{\tau_{0}}$-dependence as
well as the presence of a weighting factor describing the effect of waveguide
coupling $K\equiv\frac{\tau_{ex}}{\tau_{0}}$ and pump detuning $\Delta\omega$.
The optical-Q scaling dependences fall into two regimes. The first occurs when
$\omega_{m}<\frac{1}{\tau}$ . In this regime the mechanical oscillation
threshold exhibits an inverse cubic dependence on optical Q ($P\propto\frac
{1}{Q_{m}}\left(  \frac{1}{Q_{0}}\right)  ^{3}$) . In contrast, for
$\omega_{m}>\frac{1}{\tau}$ (herein called the high-frequency (HF) regime),
the rapid $1/Q_{0}^{3}$ dependence is reduced because the Stokes field
build-up is less-and-less effective in creating radiation pressure. In this
regime, minimum threshold can be shown to occur over-coupled (i.e., $K>1$),
where again the condition $\omega_{m}<\frac{1}{\tau}$ is met (i.e. the
mechanical oscillation frequency is again less than the "loaded" cavity
bandwidth), which causes the minimum threshold (i.e., $\frac{\partial^{2}%
P}{\partial K\partial\Delta\omega}=0$ \cite{thresholdOptimization}) to
approach an \textit{asymptotic} \textit{value}. The transition to the HF
regime, under conditions of optimum threshold, occurs for an optical Q given
by $Q_{0}^{HF}\approx\frac{\omega_{0}}{\omega_{m}}$.

To confirm these theoretical predictions the threshold dependence (as given by
Eq. 2) on both optical and mechanical Q-factor have been measured. The data
presented are taken using a single microtoroid device. Coupling to the
resonator was performed using a fiber-optic taper coupler (see inset fig.2).
The micro-toroid under consideration had principal, pillar and minor diameters
of $72,36$ and $6.8\mu m$, respectively, and possessed mechanical resonances
frequencies at $4.4$, $25.6$ and $49.8$ MHz for the first three mechanical
modes ($n=1,2,3$ and $m=0$). The optical pump wavelength was $\sim$1550 nm and
mechanical oscillation instability was observed by detecting the
characteristic oscillations in the transmitted pump power (compare Fig. 1c)
\cite{Rokhsari} using an electrical spectrum analyzer (ESA) as described in
Ref \cite{Rokhsari}. Optimization of coupling ($K$) was performed by
adjustment of the gap between the fiber taper and the microtoroid as described
in refs. \cite{Spillane,Kippenberg4}. To measure the dependence of the
oscillation threshold on $Q_{m}$, a silica microprobe was brought into contact
with the interior (disk region) of the toroid structure. Variation of the
probe contact force thereby modified mechanical Q while leaving the optical Q
unaffected. The microprobe, which was made from an optical fiber, had a tip
diameter of $\sim$2 $\mu m$ and can be seen in the inset of Fig. 2. In the
absence of probe contact $Q_{m}$ was measured to be $\sim5000$ for the $n=1$
mode, and upon progressive increase in tip pressure could be continuously
decreased to $Q_{m}\approx$ $50$. Below threshold, the thermal displacement of
the mechanical eigenmodes (the temperature being $300$K) provides sufficient
modulation to be optically detectable, causing the appearance of Lorenzian
peaks in the cavity transmission spectrum. $Q_{m}$ was then determined by
fitting the transmission spectrum with a Lorenzian, as shown in the inset of
Fig. 2. For each $Q_{m}$, the minimum threshold was measured for the $n=1$
flexural mode as shown in Fig. 2. The solid line in the main panel shows that
the data exhibit the $1/Q_{m}$ dependence in agreement with Eqn.
\ref{EqThreshold}, and Ref. \cite{Braginsky4}.

We next measured the threshold dependence on the optical Q factor as shown in
Fig. 3 for both the $n=1$ (main panel) and the $n=3$ (inset) mechanical modes.
The optical Q factor was adjusted by exciting different radial and transverse
optical modes. For lower optical Q, wherein the acoustical oscillation
frequency falls within the cavity bandwidth, the rapid $1/Q^{3}$ dependence is
observed for $n=1$ as predicted. For higher optical Q, as theoretically
predicted a transition into the HF regime occurs at $Q_{0}^{HF}\approx10^{7}%
$.\ This point agrees well with the theoretical prediction ($\frac{\omega_{0}%
}{\omega_{m}}$). It is important to note that these observations rule out
thermal effects \cite{Zalalutdinov2} as origin of the observed
oscillations\cite{thermal}. In Fig. 3, the solid line is the minimum threshold
i.e. ($\frac{\partial^{2}P}{\partial K\partial\Delta\omega}=0$) as given by
equation (2). With the exception of the effective mass, $m_{eff}$, all
parameters used to create this plot were experimentally measured parameters
(i.e., $C(\Gamma),R,Q_{m},\omega_{m},Q_{0},\omega_{0}) $. The effective mass
$m_{eff}$ was inferred to be $m_{eff}^{(1)}$ $=3.3\times10^{-8}$ $kg$.

The inset of Fig. 3 shows the measured threshold versus Q for the $n=3$ mode.
The $n=3$ mode threshold dependence shows that this mode is already well into
the HF regime, exhibiting the\ theoretically predicted asymptotic behavior of
the minimum threshold. This fact is consistent with the observed resonance
frequency, 49 MHz, for the $n=3$ mode which predicts that the HF regime occurs
for optical Q factors in excess of 10$^{7}$ ($Q_{0}^{HF}=\frac{\omega_{0}%
}{\omega_{m}}=3.8\times10^{6}$). Comparison with the $n=1$ mode data shows
that oscillation on the $n=3$ mode is preferred for lower optical Qs. Indeed,
preference to the $n=3$ mode was possible by loading the microcavity into the
over-coupled regime, in agreement with theory. The solid curve in the inset
gives the single-parameter fit to the $n=3$ data which yields $m_{eff}%
^{(3)}=5\times10^{-11}kg$, which is a factor of 660 lower than the mass of the
$n=1$ mode.

As a further test of the validity of the theoretical model, the
\textit{experimental} effective mass values are compared with the
\textit{theoretical }prediction based on finite element modeling. For the
$n=3$ mode, the predicted mass associated with the radial motion was
$m_{eff}^{(3)}=5\times10^{-11}kg$, which is in very good agreement with the
experimental fit. However, for the $n=1,2$ modes, the calculated effective
mass is a strong function of the offset of the toroidal ring with respect to
the equatorial plane of the disk\cite{Doubling}. To both validate and quantify
this offset, a cross section of the toroid microcavity used in this study was
obtained with focused ion beam slicing. SEM imaging (cf. Fig. 4 panel A)
reveals the presence of the above-postulated equatorial offset which amounts
to $\Delta=1.3$ $\mu m$ . Incorporation of this offset into the numerical mass
calculation yields $m_{eff}^{(1)}=2.6\times10^{-8}kg$ and $m_{eff}%
^{(2)}=2\times10^{-9}kg$. This value agrees very well with the experimental
values from above. Finally, the numerical model also explains why the $n=2$
mode is only observed sub-threshold in the experiments. The low mechanical Q
value ( $\sim200$) in conjunction with its high effective mass and frequency,
predicts threshold powers $>2$ mW, which are higher than pump powers available
in the experiments.

In summary, presented is both an experimental and theoretical analysis of
radiation pressure induced parametric oscillation instability, as predicted by
Braginsky. Excellent agreement of the threshold functional dependence on
optical Q is obtained, providing a confirmation that radiation pressure is the
excitation mechanism of the observed oscillations. Besides the fundamental
aspects of this work, the observed coupling of mechanical and optical modes by
radiation pressure can find applications in micro- and nanomechanical systems
(MEMS/NEMS)\cite{Craighead} for ultra-high sensitivity measurements of charge
\cite{Cleland2}, displacement \cite{Rugar2}, mass, force\cite{Rugar2} or
biological entities \cite{Ilic}. Equally important, radiation pressure as
observed here can be used to achieve cooling of a mechanical resonator mode.

This work was supported by the NSF, DARPA and the Caltech Lee Center. T.J.K.
acknowledges an IST-CPI postdoctoral fellowship.

\newpage

\begin{figure}[ptb]
\begin{center}
\includegraphics[width=3.5in
]{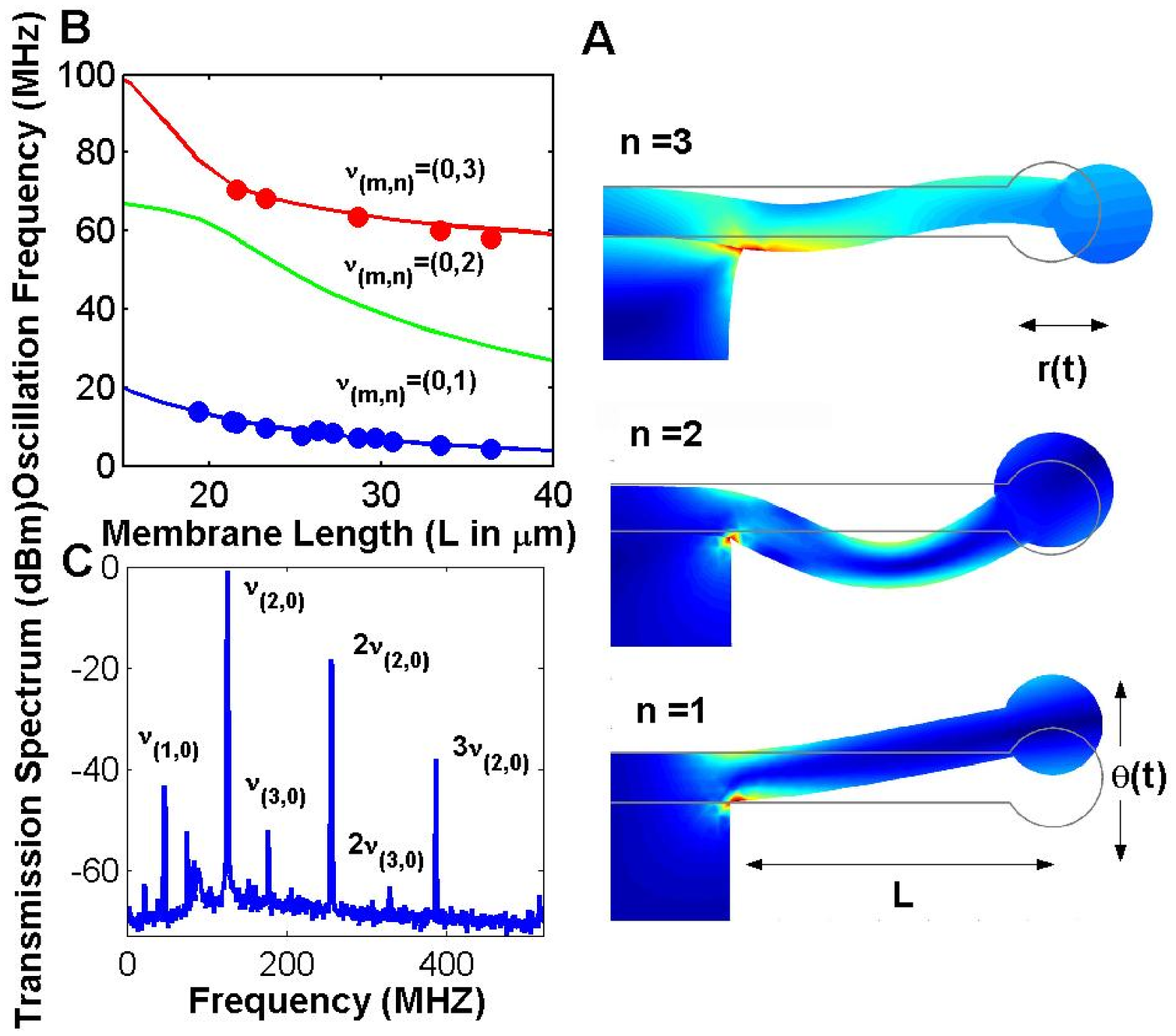}
\end{center}
\caption{ Panel A: Finite element modeling of the micro-mechanical resonances
of a silica toroid microcavity (diameter of the toroid periphery is 4 $\mu
m$). The radial mode and azimuthal mode order are denoted with $n$ and
$m,$(where $m=0$ corresponds to rotationally symmetric modes). Shown are the
first three (rotationally symmetric radial modes ($n=1,2,3$ $m=0$) in cross
section with amplitude of motion greatly exaggerated for clarity. In addition
the stress field is indicated using color. Note that the mechanical motion
modulates the cavity pathlength due to a change in the cavity radius, which
causes the excitation of optical sidebands. Panel B: Mechanical oscillation
frequency versus the cavity length $L$ for the first three fundamental
mechanical modes (Dots are experimentally measured frequencies from Ref.
\cite{Rokhsari}). Panel C: Typical frequency spectrum of the optical cavity
transmission, revealing the presence of several sidebands which correspond to
the first three mechanical eigenmodes ($\omega_{m})$, as well as harmonics of
$\omega_{m}$. Here, only the $n=2$ mode is above threshold, whereas the
remaining mechanical modes are subthreshold\ (and observable due to their room
temperature thermal displacement noise).}%
\end{figure}\newpage\begin{figure}[ptbptb]
\begin{center}
\includegraphics[width=3.5in
]{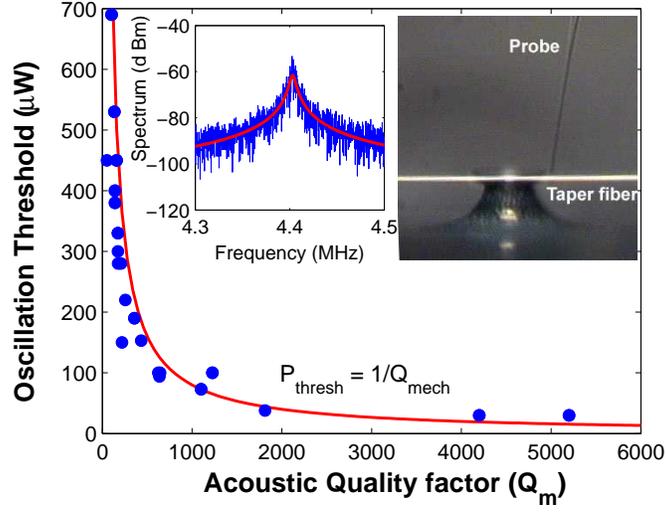}
\end{center}
\caption{Main panel: The oscillation threshold ($\mu Watts$) versus the
mechanical quality factor of the $n=1$ mode. The solid line is the theoretical
prediction based on an inverse Q relation ($P\propto1/Q_{m}$). Right inset:
Side-view, optical micrograph of the experimental setup, consisting of a
silica microprobe in contact with a taper-fiber-coupled micro-toroid of
72-$\mu m$-principal diameter. Left inset: Spectrum of the optical cavity
transmission exhibiting the thermal displacement noise of the $n=1$ mechanical
mode. Solid line: The Lorenzian fit to infer the value of $Q_{m}$.}%
\end{figure}

\begin{figure}[ptb]
\begin{center}
\includegraphics[width=3.5in
]{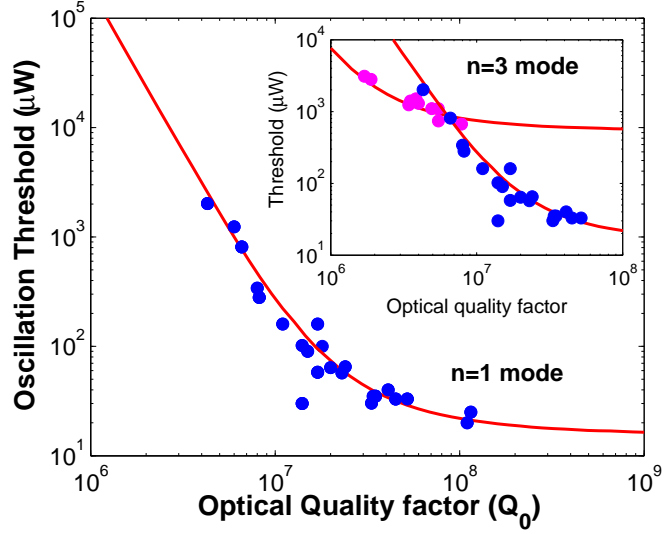}
\end{center}
\caption{Main panel: The measured mechanical oscillation threshold ($\mu
Watts$) plotted versus the optical Q factor for the fundamental flexural mode
$(n=1,\nu_{m}=4.4$ MHz$,Q_{m}=3500)$. The solid line is a one-parameter
theoretical fit obtained from the minimum threshold equation by first
performing a minimization with respect to coupling and pump wavelength
detuning, and then fitting by adjustment of the effective mass (\ $m_{eff}%
^{(1)}$ $=6.6\times10^{-8}$ $kg$). Inset: The measured threshold for the
3$^{rd}$order mode $(n=3,\nu_{m}=49$ MHz$,Q=2500)$ plotted versus optical Q.
The solid line gives again the theoretical prediction with $m_{eff}%
^{(3)}=1.1\times10^{-10}$ $kg$. The $n=1$ mode data from the main panel is
superimposed for comparison.}%
\end{figure}

\begin{figure}[ptb]
\begin{center}
\includegraphics[width=3.5in
]{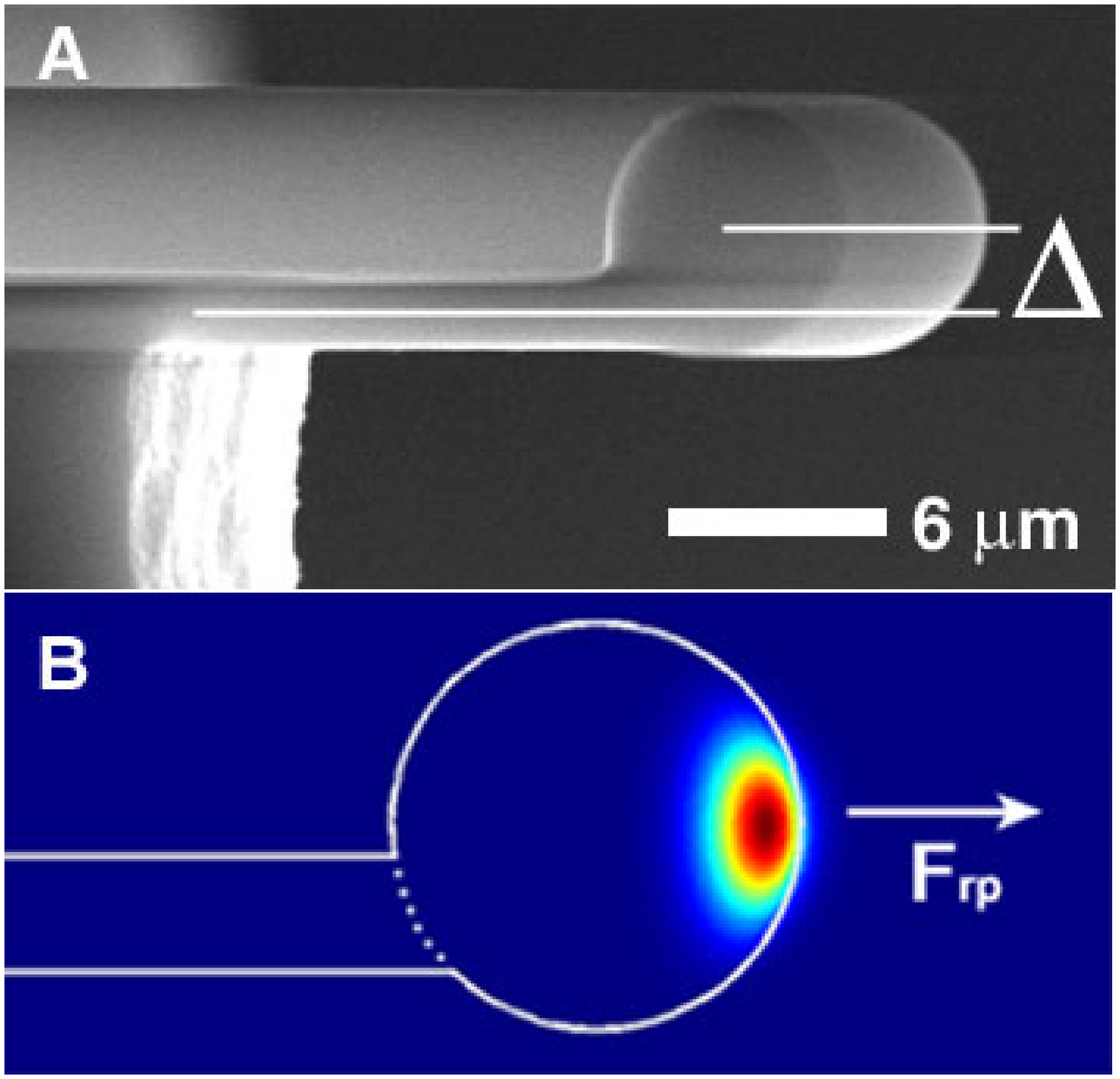}
\end{center}
\caption{Panels A: Scanning-electron micrograph of the microcavity cross
section achieved using focused-ion beam (FIB) preparation. The image reveals
the presence of an offset in the toroid with respect to the 2-$\mu m$ thick
silica support disk (offset of ca 1.3 $\mu m$). Panel B: Finite element
modeling of the fundamental optical mode. The presence of an offset provides a
moment-arm that effectively enhances coupling of radiation pressure to the
$n=1$ transverse motion.}%
\end{figure}


\begin{thebibliography}{20}
\bibitem[{Str()}]{StressOpticalEffect}
\bibinfo{note}{From the simulations it is calculated that the change in
  refractive index due to the stress-optical effect is more than one order of
  magnitude smaller compared to the direct spatial change in cavity path-length
  (i.e. ${2\pi(R+dR)(\Delta n+n_{eff})\approx 2\pi n_{eff}(R+dR)) }$.}
\bibitem[{Mec()}]{MechModulation}
\bibinfo{note}{Note that the mechanical energy ($E_{m}$) is asssociated with
  both a radial motion ($\frac{1}{2}m_{eff}\omega _{m}r^{2}$) as well as a
  transverse motion. However, to highest order, only the radial motion
  modulates the cavity optical path-length so that the effective mass is given
  by calculating numerically $m_{eff}=E_{m}/(\omega _{m}^{2}r^{2})$.}
\bibitem[{the()}]{thermal}
\bibinfo{note}{Thermal instabilities are not dependent upon resonant build-up
  since the thermo-optic mechanical coupling is dependent upon the
  \textit{coupled} (i.e. absorbed) power into the cavity, not the \textit{circulating}
  power. The expected threshold dependence on optical Q for thermal
  instabilities is therefore $\frac{1}{Q_0^2}$, which is not observed in the
  current experiments.}
\bibitem[{Dou()}]{Doubling}
\bibinfo{note}{We note that such an offset must exist, since mechanical motion
  of the n=1,2 modes causes modulation of pathlength at the frequency ${2\cdot
  \omega_m}$ (i.e. frequency doubling), which is not observed in experiments}.
\end{thebibliography}
\end{document}